\documentclass[12pt,cites,a4paper]{article}
\usepackage[dvipdfmx]{graphicx}
\usepackage{color}
\usepackage{epsfig}
\usepackage{amssymb}
\usepackage{subfigure}
\usepackage{cite}
\usepackage{caption}
\begin{document}

\title{
Discrimination of time-dependent inflow properties with a cooperative dynamical system}

\author{ Hiroshi Ueno$^{a}$, Tatsuaki Tsuruyama$^{b}$, Bogdan Nowakowski$^{c}$,\\
Jerzy G\'{o}recki$^{c,*}$ and Kenichi Yoshikawa$^{a,*}$ \\
$^{a}$ Faculty of Life and Medical Sciences,\\
Doshisha University, Kyoto 610-0394, Japan\\
$^{b}$Department of Diagnostic Pathology, Graduate School of Medicine,\\
Kyoto University, Japan\\
$^{c}$Institute of Physical Chemistry, Polish Academy of Sciences,\\
Kasprzaka 44/52, 01-224 Warsaw, Poland.\\
$^{*}$jgorecki@ichf.edu.pl\\
$^{*}$keyoshik@mail.doshisha.ac.jp\\
}

\date{\today}
\maketitle

\begin{abstract}
% insert abstract here
\noindent
Many physical, chemical and biological systems exhibit a cooperative or sigmoidal response with respect to the input. In biochemistry, such behavior is called an allosteric effect. Here we demonstrate that a system with such properties can be used to discriminate the amplitude or frequency of an external periodic perturbation or input. Numerical simulations performed for a model sigmoidal kinetics illustrate that there exists a narrow range of frequencies and amplitudes within which the system evolves toward significantly different states. Therefore, observation of system evolution should provide information about the characteristics of the perturbation. The discrimination properties for periodic perturbation are generic. They can be observed in various dynamical systems and for different types of periodic perturbation. 
\end{abstract}
%\pacs{ PACS: ?}

\section{Introduction}

 Bistability and hysteresis are commonly observed in physics, chemistry and biology \cite{murr,rus,KY1, SMI1, TK1}.
Let us assume that a system has two stable states $S_1$ and $S_2$ and that an increase in the value of control parameter $\lambda$ above the threshold $\lambda_1$ triggers the transition from $S_1$ to $S_2$, whereas the reverse transition from $S_2$ to $S_1$ occurs if the value of the control parameter drops below $\lambda_2$. Such a system can obviously be used as a discriminator of the control parameter value. For example, if the initial state is $S_1$ and after some time we observe the system in $S_2$, then at some point the value of the control parameter necessarily exceeded $\lambda_1$. However, if only time-monotonic changes in the value of the control parameter are considered, then the system discrimination ability is reduced to just two values ; $\lambda_1$ and $\lambda_2$.

In this paper we demonstrate the suitability of a dynamical system characterized by sigmoidal kinetics for discrimination-oriented applications, under a new strategy of imposing a periodic perturbation or input on a cooperative system. It has been reported \cite{kaw1,kaw2,kaw3} that periodic perturbations can significantly change the time evolution of a nonlinear system. As a discriminator prototype, we consider a two-variable system in which the inflow of one of the variables is a control parameter. Numerical simulations reveal a non-trivial property of such a system: a marginal change in the inflow parameters (amplitude or frequency) can switch the response of a cooperative system between different branches in the stage diagram. The frequency at which such switching occurs is a monotonic function of the inflow amplitude. Therefore, at a fixed amplitude of periodic inflow, the observation of a transition between different types of oscillatory evolution of the system provides information which allows us to discriminate the inflow frequency. The above discussion does not necessarily limit the range of frequencies that can be discriminated by the observation of transitions between different types of oscillations Similarly, for a fixed frequency of periodic inflow, transitions between different types of system oscillations occur within a narrow range of amplitudes. The transition can be used to discriminate the inflow amplitude, but for the model considered here, the useful range of such discrimination is rather limited.

In numerical simulations, we consider simple system dynamics defined by a single sigmoidal term expressed by a rational function, which is typical for enzymatic reactions \cite{hill,dr1,dr2,dr3,dr4}. In such reactions the appearance
of sigmoidal kinetic behavior is usually interpreted to be the result of the interaction
of substrates with enzymes through positive cooperative binding. Modeling
 of cooperative binding leads to the Hill equation \cite{hill}: 
\begin{equation}
\theta = \frac{[ L]^n}{K_d + [ L]^n}
\label{hill}
\end{equation}
where $\theta$ is the fraction of ligand binding sites filled, $[L]$ is the ligand concentration, $K_d$ is the apparent dissociation constant derived from the mass action
law, and $n$ is the Hill coefficient which represents the degree of cooperativity. 
If $n = 1$, there is no cooperativity; for $n > 1$, the cooperativity is positive. Kinetics with sigmoidal behavior are not limited to enzymatic reactions. This also describes the response of various biological systems to external stimuli, including the effect of drug delivery, which is an interesting topic in pharmacology. Among the many experimental studies that have reported sigmoidal behavior, the Hill coefficient $n$ usually has a value between 2 and 4 \cite{hill,dr1,dr2,dr3,dr4,MAPK1, MAPK2,MAPK3,MAPK4,tsuru1, tsuru2}. Here we selected $n$ = 3 for the numerical simulations presented below. 

The paper is organized as follows. In the next section, we consider a bistable model and study its time evolution as a function of the amplitude and frequency of periodic inflow. We demonstrate how the system can be used as a discriminator and discuss the sources of discrimination errors. In the final section we argue that the observed phenomenon is generic and discuss its potential applications.

\clearpage

\section{The response of a model dynamical system to periodic perturbations}

%§ definition of dynamical system
Let us consider a dynamical system of two variables $(x(t)$, $y(t))$ defined by a set of differential equations:
\begin{equation}
\frac{d x}{d t}=g(x,y,t) =
 -\alpha x + y + A\cdot (\sin(2\pi f t + \phi_0) + 1) \cdot \Theta(t)
\label{r1}
\end{equation}
\begin{equation}
\frac{d y}{d t}=h(x,y) =
\frac{1}{\varepsilon} \cdot (\frac{x^3}{1+x^3} - y)
\label{r2}
\end{equation}
%say something on phase / frequency
In Eq.(\ref{r1}), the last term $ I(t) = A\cdot (\sin(2\pi f t + \phi_0) + 1) \cdot \Theta(t)$ describes a periodic inflow of $x$ with frequency $f$ and initial (for $t=0$) phase $\phi_0$. $\Theta(t)$ is the Heaviside step function. We assume that the there is no inflow for $t<0$, and it is switched on at $t=0$. If $\phi_0 = 3 \cdot \pi /2$, then $I(t)$ is a continuous function. In this case, $I(t=0) = 0$. It then increases and finally oscillates. For any other phase, the inflow term is not continuous at $t=0$; for example, if $\phi_0 = \pi /2$ and then $I(t=0) = 2\cdot A$. $I(t)$ then decreases and finally oscillates. The inflow term is always non-negative. For $t > 0$ the time average of $ I(t)$ equals $A$ and is independent of the frequency and the initial phase. If the inflow amplitude $A = 0$, then $(x = 0, y= 0)$ is the only steady state of Eqs.(\ref{r1},\ref{r2}) and is stable. In the following analysis, we assume that the stable state of the system without flow is the initial state for the simulated evolution.

\begin{figure}[!tbp]
\includegraphics[angle=0,width=0.9\textwidth]{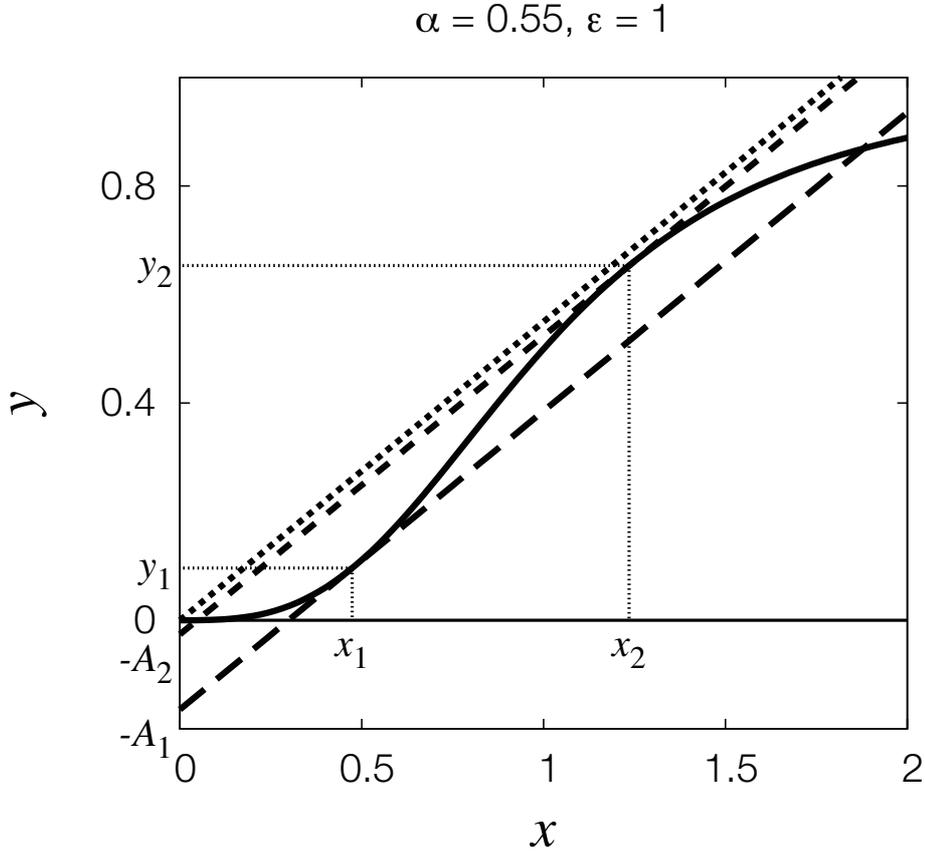}\\
\caption{ Positions of nullclines for the dynamical system defined by Eqs.(\ref{r1},\ref{r2}). The model parameters are: $\varepsilon = 1$, $\alpha = 0.55$, $f = 0$ and $\phi_0 = 0$. The nullcline $h(x, y) = 0$ is plotted with a solid line. The nullcline $g(x, y,t) = 0$ is shown for a few cases: $A = 0$ ( dotted line), $A = A_2=0.02603 $ ( short-dashed line), $A = A_1 = 0.16445$ ( long-dashed line). For the selected parameters of the model, the variables at tangential points are $(x_1,y_1)=(0.47368,0.09607)$ and $(x_2,y_2)=(1.23531,0.65339)$ }
\label{fig1}
\end{figure}

 Initially, let us consider the time evolution of the system for a constant inflow $ I(t) = A >0$ for $ t \ge 0$ (thus, $f = 0$ and $\phi_0 = 0$). The characteristics of the time evolution depend on the amplitude of the inflow term and on the initial state. In this case, the nullcline $g(x, y, t) = 0$ is the time-independent line with a definite slope determined by the value of $\alpha$ and a shift which depends on the inflow amplitude $A$. Figure \ref{fig1} shows the location of nullclines, calculated for $\varepsilon = 1$, $\alpha = 0.55$ and a few different values of the inflow. Let us assume that $A_1$ and $A_2$ are the amplitudes for which the BN $g(x, y) = 0$ nullcline is tangential to the sigmoidal-shaped nullcline $h(x, y) = 0$. The stable stationary states of the system can be located on two branches on the $h(x, y) = 0$ nullcline. One contains all of the points of the $h(x, y) = 0$ nullcline located between point $(0,0)$ and the tangency point $(x_1, y_1)$. We will call it the lower stable branch (LSB). The other is the upper stable branch (USB), and is formed by all of the points of the $h(x, y) = 0$ nullcline located above $(x_2, y_2)$. The stationary states located on the nullcline between $(x_1, y_1)$ and $(x_2, y_2)$ are unstable. In the case when $A < A_2$, the only stationary state is located on the lower stable branch, so the system converges to the stable state $ y_\infty = \lim_{t \to \infty} y(t)$ such that $ y_\infty \le y_1$ regardless of the initial state. Similarly, for $A > A_1$, the single BN stationary state is located on the upper stable branch, and for all initial states the system converges to the stable state $ y_\infty \ge y_2$. For $A_2 \le A \le A_1$, the stationary state that is approached for ${t \to \infty}$ depends on the initial state and on the partition of the phase space determined by the separatrices of the saddle point which is located on the middle branch of the $h$ nullcline. This analysis also applies when the frequency of inflow oscillations is very high. In such a case, the flow oscillations are much faster than both the system dynamics and the system responses to the time-averaged value of the inflow $A$.

For sufficiently slow oscillations of the inflow ( $0 < f \ll 1$), the system can follow the slowly relocating stable state, the position of which varies according to the instantaneous value of the inflow. If the initial state of the system is $(x(0) = 0, y(0) = 0)$ and $2 \cdot A \le A_1$, then $ y(t) \le y_1$ for all $t$. Therefore, the system state oscillates along the lower stable branch of the $h(x, y)=0$ nullcline with the period defined by the frequency of inflow oscillations. If $2\cdot A > A_1$, then there are intervals of time within which the system has a single stationary state located on the upper stable branch. During a single oscillation cycle, there are moments of time $t_1$ and $t_2$ at which $y(t_1) \le y_1$ and $ y(t_2) \ge y_2$, and thus oscillations that extend over both stable branches are expected.

 For moderate values of $f$, the system dynamics are too slow to closely follow the changes in the inflow value. In such a case, oscillations around a stable state located in the lower stable branch that extend to the unstable branch, as well as oscillations around a stable state located in the upper stable branch that extend to the unstable branch, should be observed. This is confirmed by numerical simulations. 

 The complexity of oscillations observed for a constant inflow amplitude $A=0.12$ (thus $ A > A_1 / 2$ but $ A < A_1$ ) as a function of flow frequency is illustrated in Fig. 2. As discussed above, for the selected amplitude and a low frequency of inflow oscillations, the system dynamics follow the time-dependent stationary state and oscillations of $y(t)$ extend over both stable branches of the $h(x, y)=0$ nullcline. For intermediate frequencies, oscillations accumulate on the upper stable branch of the nullcline and the minimum value of $y(t)$ increases with frequency. Then, at a certain frequency $f_{c}$ ( for the selected amplitude of oscillations, $f_{c} \approx 0.0312$ ), the oscillations switch from the upper to the lower stable branch of the $h(x, y)=0$ nullcline. The transition between oscillations located on different stable branches is quite pronounced and should be easily detected in experiments with a system exhibiting hysteresis. Therefore, it becomes apparent that a cooperative system can discriminate the frequency of a perturbation if its amplitude remains fixed. Numerical simulations have also demonstrated that the frequency of the transition between oscillations on the USB and LSB depends on the phase $\phi_0 $. The right upper corner of Fig. 2 shows two types of oscillations that are observed for $f=0.031$. If $\phi_{0}=0$, the system oscillates at the upper stable branch, but if $\phi_{0}=\frac{3\pi}{2}$, oscillations around the lower stable branch are seen. Fortunately for the application of this approach to discrimination, the interval of frequencies within which phase-dependent evolution is observed is very narrow. For $A=0.12$, it is $[0.0306, 0.0312]$. The width of this interval ($\Delta f \sim 0.0006$) defines the precision in frequency discrimination.

The dynamical system considered here can also be used to discriminate the amplitude of an applied perturbation. Figure 3 shows the time evolution of $y(t)$ for a few values of perturbation amplitude $A$ and a fixed inflow frequency ($f=0.05$). As expected, for small amplitudes ($A \le 0.1122$) the oscillations of $y(t)$ are limited to the LSB. For larger amplitudes ($ 0.1122 < A < 0.1361$), the range of observed values of $y(t)$ increases, but the oscillations are still anchored on the LSB. Finally, if $ 0.13617 \le A$, the oscillations move onto the USB. The transition between the different types of oscillation is quite pronounced and can be used to discriminate the value of amplitude. Here, similar to the cases illustrated in Fig. 2, we observe a narrow interval of amplitudes ( $\Delta A \sim 0.0001$) within which the type of oscillation depends on the initial phase.

\clearpage

\clearpage

\begin{figure}[!tbp]
\includegraphics[angle=0,width=0.9\textwidth]{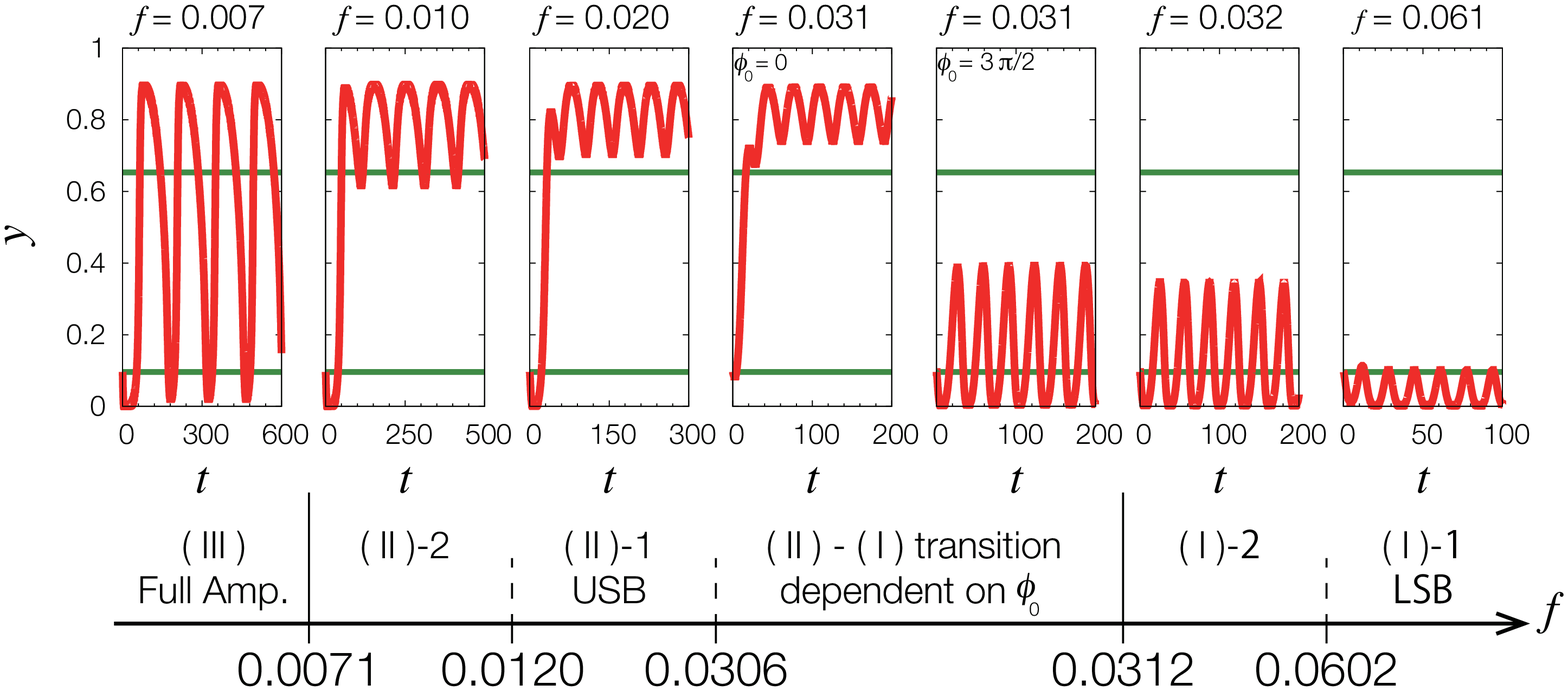}\\
\caption{ Time evolution of the dynamical system described by Eqs.(\ref{r1},\ref{r2}) as a function of the inflow frequency $f$ for a fixed amplitude $A=0.12$. The model parameters are: $\varepsilon = 1$, $\alpha = 0.55$. Tics and numbers on the frequency scale mark transitions between different types of oscillation. The initial phase is $ \phi_0 = 3 \pi /2$ for all cases except the central one, for which $ \phi_0 = 0$. The horizontal green  lines mark the values of $y_1$ and $y_2$.}
\label{fig2}
\end{figure}

\clearpage

\begin{figure}[!tbp]
\includegraphics[angle=0,width=0.9\textwidth]{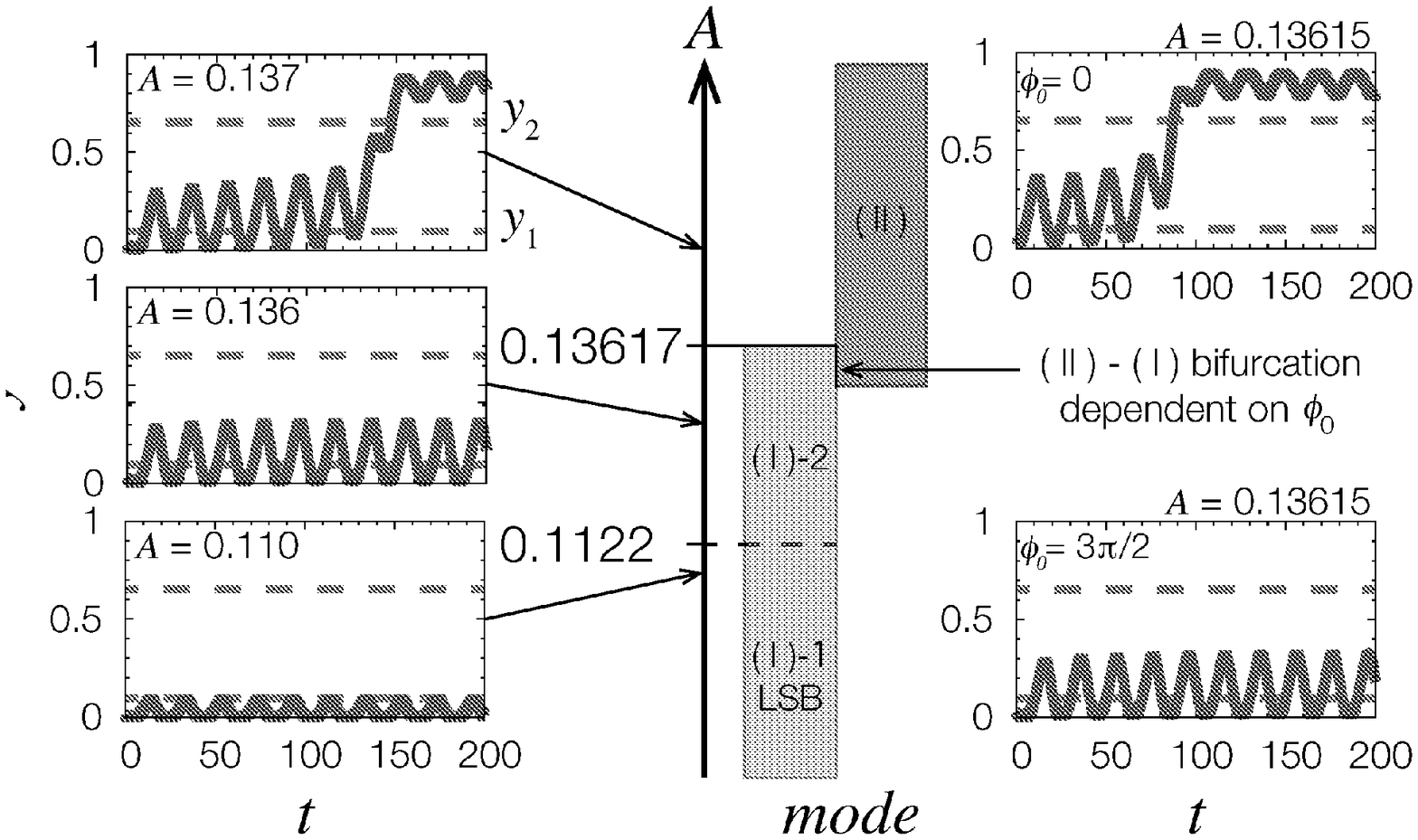}\\
\caption{ Time evolution of the dynamical system described by Eqs.(\ref{r1},\ref{r2}) as a function of the inflow amplitude $A$ for a fixed frequency $f=0.05$. Tics and numbers on the frequency scale mark transitions between different types of oscillation. The initial phase is $ \phi_0 = 3 \pi /2$ for all cases except at the top in the right column, for which $ \phi_0 = 0$. The horizontal dashed lines mark the values of $y_1$ and $y_2$.}
\label{fig3}
\end{figure}

\clearpage

 To give a more precise description of system evolution, let us introduce a classification of oscillations based on the minimum and maximum values of $y(t)$ observed over a long time interval for which the evolution has reached a stationary state. We define:

\begin{equation}
y_{min} =
min_{t \in \left[t_{min}, t_{max} \right]} \hskip 0.2cm y(t)
\label{y_min}
\end{equation}
and
\begin{equation}
y_{max} =
max _{t \in \left[t_{min}, t_{max} \right]} \hskip 0.2cm y(t)
\label{y_min}
\end{equation}
 Initially, we used $t_{min} = 1000$, where $ t_{max}=t_{min}+ 1000$. Next we repeated the calculations for $t_{min} = 2000$. If there is a significant discrepancy in $y_{min}$, $y_{max}$ obtained for these time intervals, then the procedure is repeated with $t_{min}$ increased by an additional $1000$ time units until agreement is attained.

The type of oscillation is classified through the comparison of $y_{min}$ and $y_{max}$ with the values of $y_1$ and $y_2$, as illustrated in Fig. 4. The classification of oscillations is summarized in Table I.

\vskip 1cm
Table I
\vskip 1cm
\begin{tabular}{ l | c }
 oscillation class & condition \\
 (I)-1 LSB & $y_{min} \le y_{max}< y_{1}$ (oscillations limited to LSB) \\
 (I)-2 & $y_{min}<y_{1}, y_{1}\le y_{max}< y_{2}$\\
 (III) & $y_{min}<y_{1}, y_{max}\ge y_{2}$\\
 (II)-2 & $y_{1}\le y_{min}< y_{2}, y_{max}\ge y_{2}$\\
 (II)-1 USB & $y_{max} \ge y_{min}> y_{2}$ (oscillations limited to USB)\\
 (IV) & $y_{1}\le y_{min}< y_{2}, y_{1}\le y_{max}< y_{2}$\\
\end{tabular}

\vskip 1cm

 As shown in Fig. 4, the transitions between oscillation types $ (I)-1 LSB \leftrightarrow (I)-2$ and $(I)-2 \leftrightarrow (III)$ are continuous because they result from an increase or decrease in the $y_{max}$ value. Similarly, the transitions between oscillation types $(III) \leftrightarrow (II)-2$ and $(II)-2 \leftrightarrow (II)-1 USB$ are continous because they are related to an increase or decrease in the $y_{min}$ value. In actual experiments, these transitions are difficult to detect because they require highly accurate data acquisition. On the other hand, the transition $(I)-2 \leftrightarrow (II)-1 USB$, on which the discrimination is based, can be easily detected because it is related to a discontinuous jump between $y_{max} < y_2$ and $y_{min} > y_2$.

\clearpage
	
\begin{figure}[!tbp]
\includegraphics[angle=0,width=0.9\textwidth]{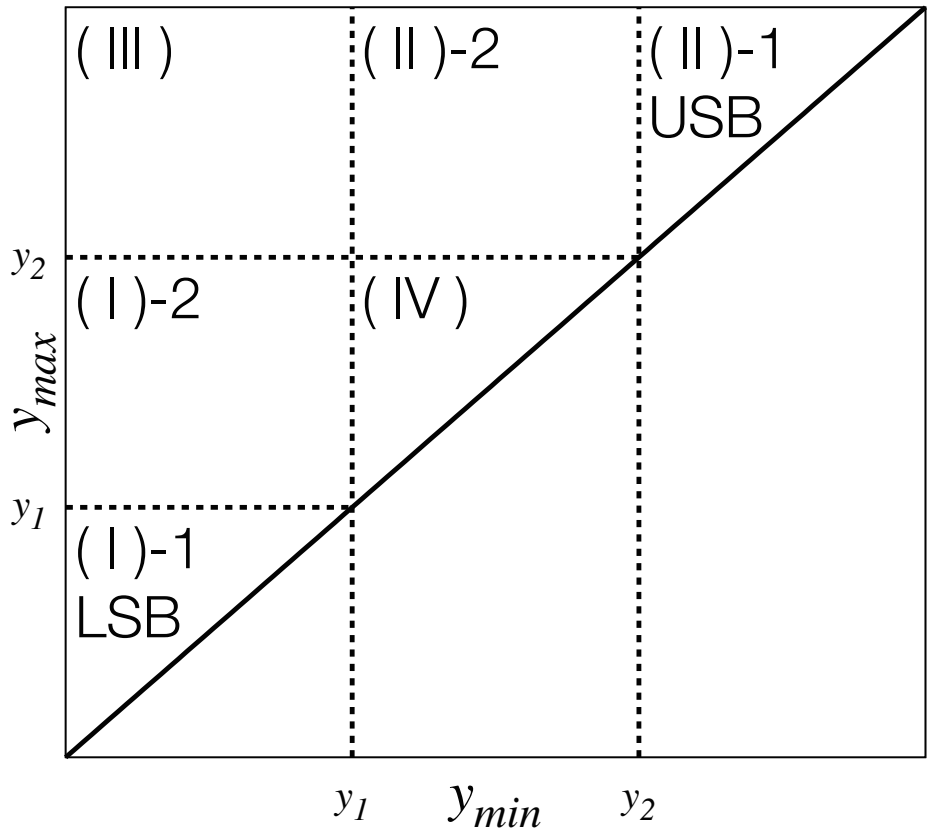}\\
\caption{ Geometrical illustration of types of oscillation in the classification based on $y_{min}$ and $y_{max}$. The dotted lines mark $y_1 = 0.09607$ and $y_2 = 0.65338$. }
\label{fig4}
\end{figure}

\clearpage

 Figure 5 illustrates the regions of parameters $(f,A)$ for which a given oscillation pattern is observed in a model characterized by $\alpha = 0.55$ and $\varepsilon = 1$. The thick line separates the region of the phase space $(f,A)$ in which class (I)-2 oscillations are observed, from the region where class (II)-1 USB oscillations appear. Let us denote points on this line as $(f_c,A_c)$. The amplitude $A_c$, when treated as a function of $f_c$, is a continuous, monotonically increasing function $A_c = G(f_c)$. Therefore, the inverse function $f_c = G^{-1}(A_c)$ exists. Our discrimination method is based on the determination of conditions in which a small change in $f_c$ or $A_c$ qualitatively changes the character of the time evolution to force a transition between $(I)-2$ and $ (II)-1 USB$ type oscillations. Let us assume that we want to measure the unknown inflow frequency and that we can regulate the inflow amplitude. The following procedure can be applied. Initially, we set a low amplitude so the system oscillates on the LSB (type $(II)-1$ oscillations). Next, the amplitude is increased up to the moment $A_z$ when oscillations of type $(II)-1 USB$ are detected. The frequency of inflow $f_z$ can be estimated as $f_z = G^{-1}(A_z)$. This method works for all frequencies greater than $f_0$, which corresponds to the tip of the $(II)-1 USB$ region ($(f_0,A_0)$). The accuracy of the estimation depends on the frequency and is high where the amplitude $A_c$ is a rapidly increasing function of $f_c$, here for $0.02 \le f_c \le 0.1$. This system can also be used to determine the amplitude of inflow when we can control the frequency. Now we set a low frequency and the system exhibits type $(III)$ oscillations. Next, the frequency is increased up to the moment $f_y$ when oscillations of type $(I)-2$ are detected. The amplitude of the inflow $A_y$ is $A_y = G(f_y)$. Unlike for frequency, the range of discriminated amplitudes does not extend outside the interval $[A_0, A_1]$.

The phase diagrams, similar to that in Fig. 5 but for $\varepsilon = 1/5$ and $\varepsilon = 5$, are shown in Fig. 6 and Fig. 7 respectively. The results are qualitatively identical to those in Fig. 5, suggesting that the described changes in the system oscillations are generic and should also apply to other systems with hysteresis influenced by a periodic perturbation.

\clearpage

\begin{figure}[!tbp]
\includegraphics[angle=0,width=0.9\textwidth]{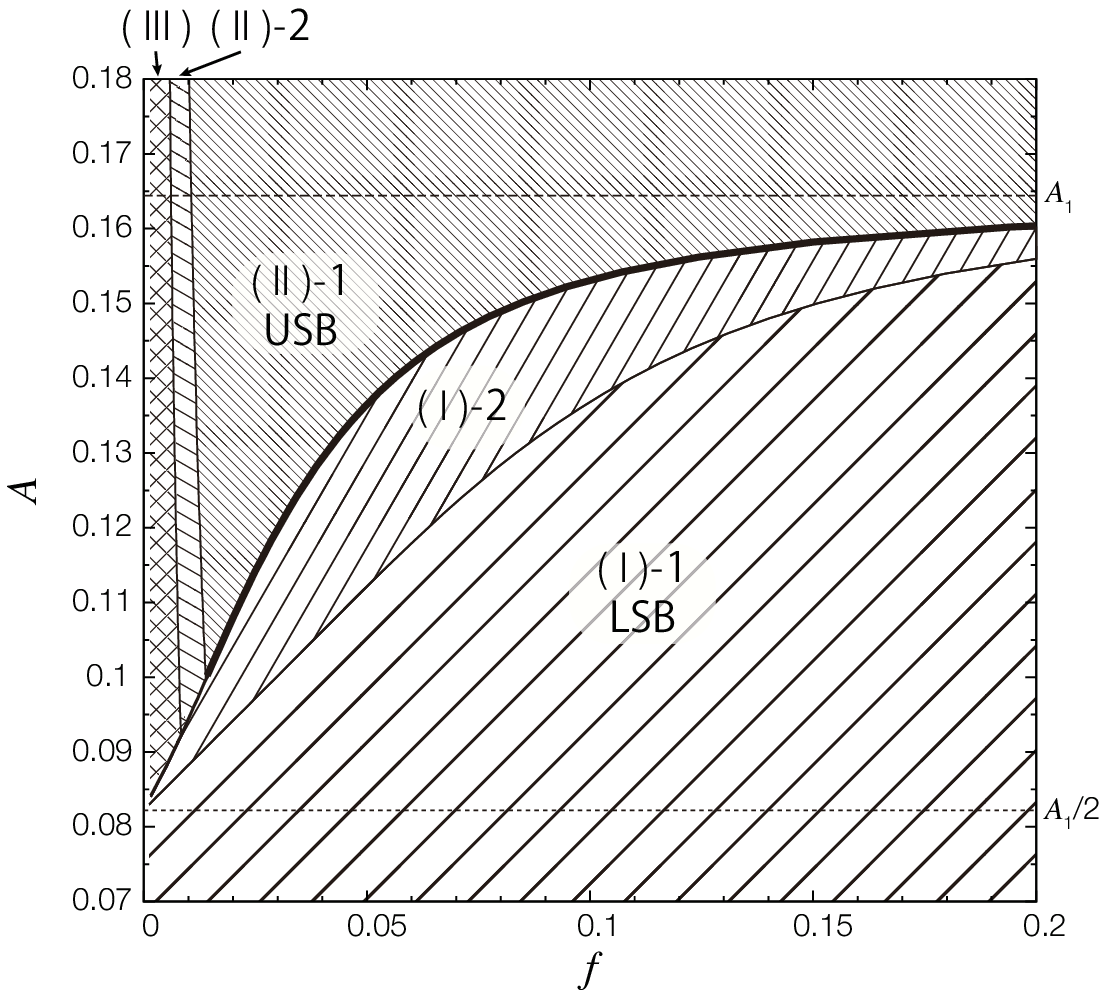}\\
\caption{ Phase diagram showing the type of oscillation as a function of inflow parameters $(f,A)$. The horizontal lines indicate $A_1 = 0.16445$ and $A_1 /2$. The model parameters are $\varepsilon = 1$ and $\alpha = 0.55$}. The thick solid line marks the boundary between oscillation classes $(I)-2$ and $ (II)-1 USB$. The transition between these oscillations is used to determine the parameters of inflow. 
\label{fig5}
\end{figure}

\clearpage

\begin{figure}[!tbp]
\includegraphics[angle=0,width=0.9\textwidth]{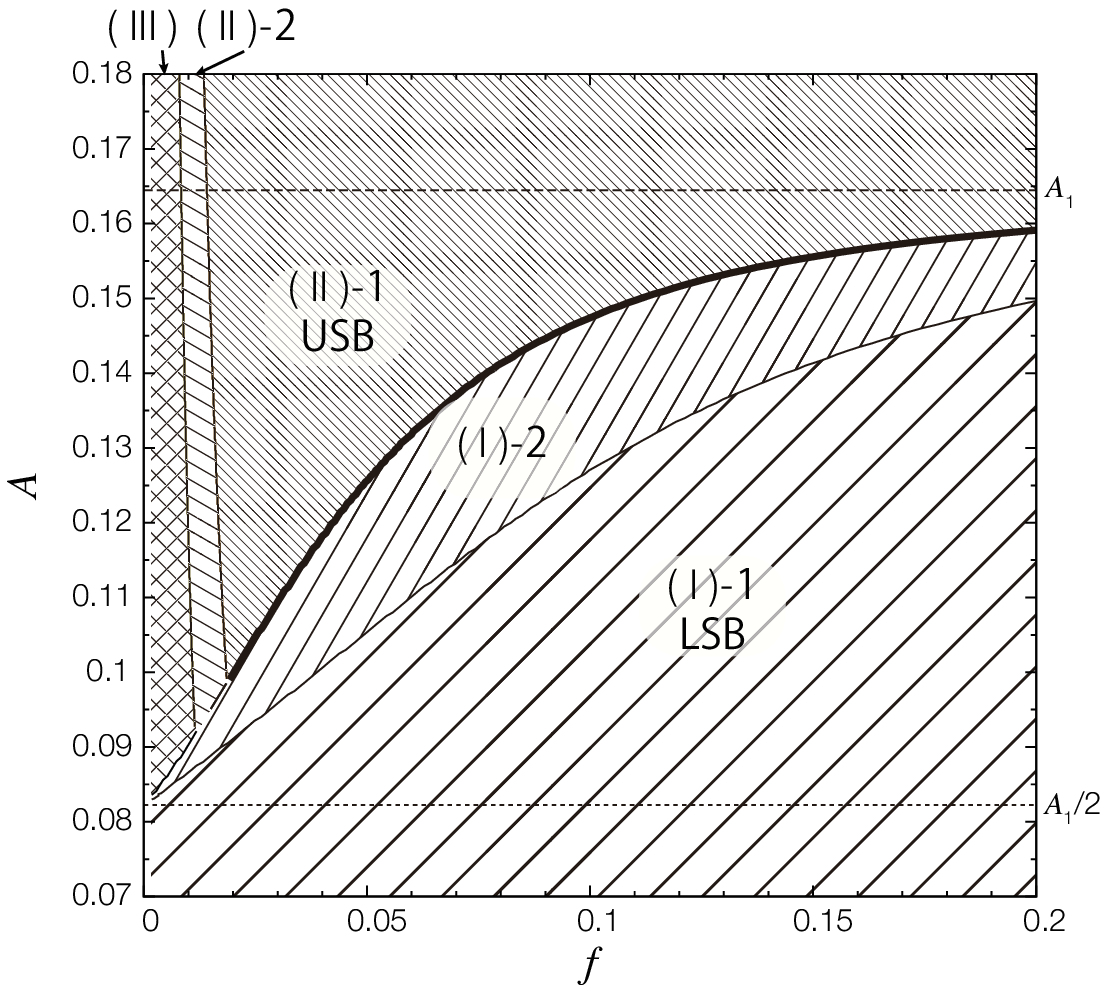}\\
\caption{ Phase diagram showing the type of oscillation
as a function of inflow parameters $(f,A)$. The horizontal lines indicate $A_1 = 0.16445$ and $A_1 /2$ . The model parameters are $\varepsilon = 1/5$ and $\alpha = 0.55$}.
\label{fig6}
\end{figure}

\clearpage

\begin{figure}[!tbp]
\includegraphics[angle=270,width=1.0\textwidth]{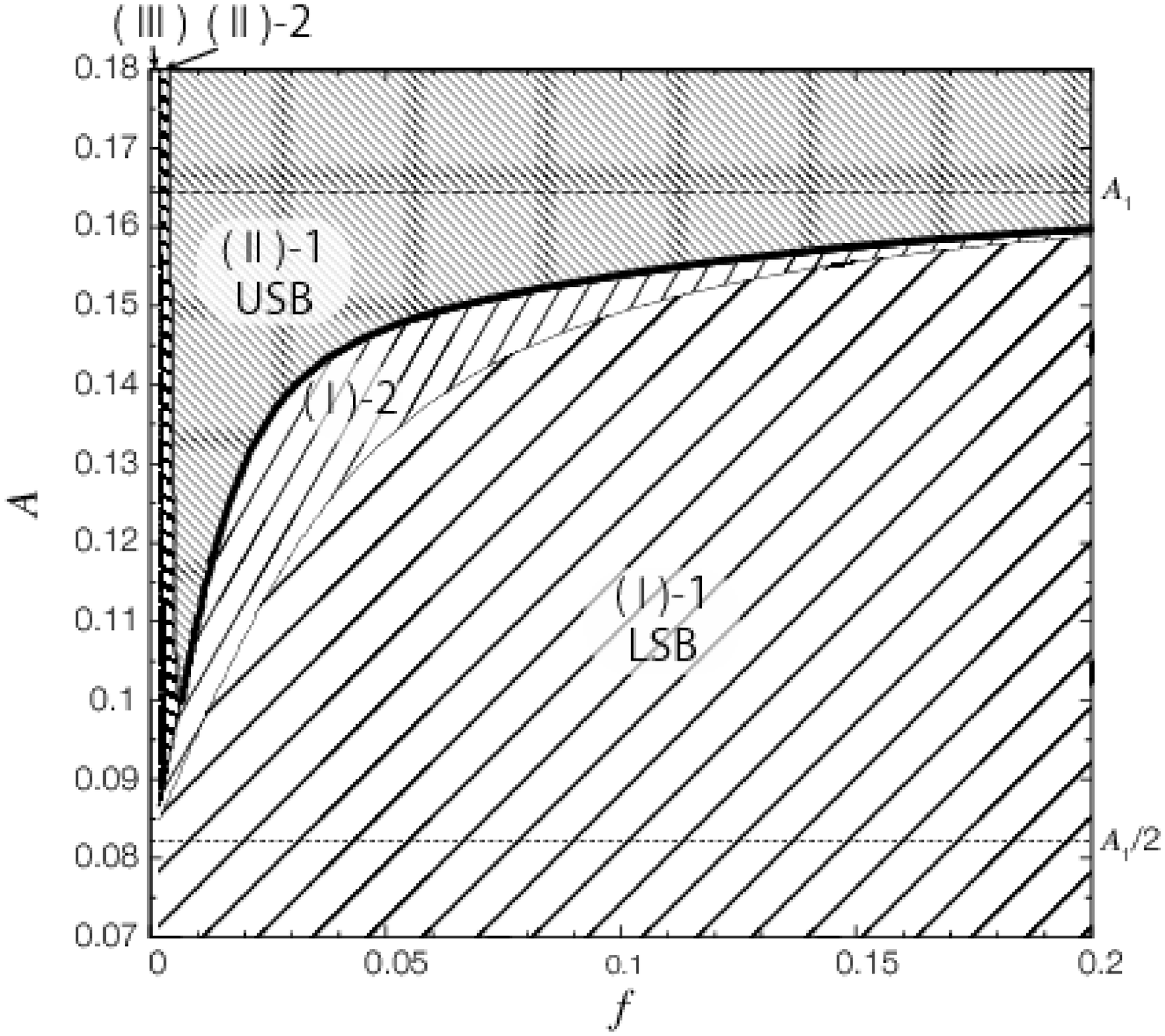}\\
\caption{ Phase diagram showing the type of oscillation as a function of inflow parameters $(f,A)$. The horizontal lines indicate $A_1 = 0.16445$ and $A_1 /2$ . The model parameters are $\varepsilon = 5$ and $\alpha = 0.55$}.
\label{fig7}
\end{figure}

\clearpage

\section{Conclusions}

 We have described how the time evolution of a cooperative system is dependent on the frequency and amplitude of a periodic stimulus. There is a narrow range of these parameters within which the characteristics of this evolution change in a qualitative manner: oscillations around one stable branch change into oscillations on another branch. This phenomenon can be used to determine the amplitude or frequency of an applied perturbation. As for the numerical framework, we evaluated the effect of the rhythmicity of substrate input in a model biochemical system with sigmoidal kinetics, i.e. $ n =3$ in the Hill equation. Using numerical simulations, we separated the phase space of inflow parameters (amplitude and frequency) into regions where specific types of oscillation are observed. The boundary line separating oscillations with significantly different behaviors (type $(I)-2$ and type $ (II)-1 USB$ oscillations) was identified. The frequency that causes a transition appears in a monotonic function of the inflow amplitude. The system can be used to determine the inflow frequency if we can control the inflow amplitude. It can also be used to determine the inflow amplitude when we can control the frequency. In other words, sigmoidal kinetics with the Hill equation can act as an inflow discriminator.

 This paper describes a system in which the nonlinear term in the kinetic equation for the $y(t)$ variable is described by $\frac{x^3}{1+x^3}$ term (cf. Eq.(\ref{r2})) and the periodic inflow is described by a trigonometric function. We believe that these results are general, and qualitatively similar behavior can be expected in other systems with cooperative characteristics. We performed numerical simulations for a model based on Eqs.( \ref{r1},\ref{r2}) but with the inflow term in the form $ J(t) = A\cdot (\tanh{(\gamma \cdot \sin(2\pi f t + \phi_0))} + 1) \cdot \Theta(t)$ for different values of $\gamma$. Such periodic inflow becomes a square-like wave for large $\gamma$. The phase diagrams that illustrate the type of oscillation as a function of $f$ and $A$ are qualitatively the same, as presented in Figs. (5-7). We also considered other nonlinear terms in the kinetic equation for $y(t)$, like $\tanh{(x-x_0)}$ or $1/({1+\exp{(- \delta \cdot(x-x_0)}})$, and obtained similar results. Therefore, we believe that real systems of chemical reactions with hysteresis can be used as discriminators in the manner described above.

The present results can be regarded as a solution to the problem of the optimum stabilization of a system in an unstable state.
Let us assume that Eqs.( \ref{r1},\ref{r2}) describe the time-dependent progress of a medical treatment where the variable $y(t)$ represents the condition of a patient. The variable $x(t)$ describes the time-dependent concentration of the curing drug. The states on the LSB and USB correspond to an ill and healthy patient, respectively. This simple model seems to realistically describe the basic features of drug therapy. It predicts that if the inflow of the drug is small, then the patient remains ill. Only a dose higher than a critical dose allows for successful treatment. However, some drugs are toxic ( such as those used in chemotherapy) and the total dose should be as small as possible. An analysis of the dynamical system presented in Fig. 5 can provide a solution: if we consider the periodic inflow of a drug in the form of Eq.(1), then the minimum amount of drug required to stabilize the patient in a healthy state corresponds to the bottom corner of the type $(II)-1 USB$ oscillation region - here $A_0 \cong 0.1$ and $f_0 \cong 0.015$.

\section{ Acknowledgement}
This work was supported by KAKENHI Grants-in-Aid for Scientific Research (15H02121, 25103012).

\par

\end{document}